\documentclass[10pt,conference]{IEEEtran}
\usepackage{amsmath,amssymb,eucal,graphicx}
\usepackage{epsfig}
\usepackage{exscale}
\usepackage{latexsym}
\usepackage{citesort}

\long\def\comment#1{}

\newfont{\bb}{msbm10 scaled 1000}
\newcommand{\CC}{\mbox{\bb C}}

\newcommand{\EE}{\mbox{\bb E}}

\newfont{\bbsmall}{msbm10 scaled 700}

\newcommand{\EEsmall}{\mbox{\bbsmall E}}

\newcommand{\ev}{{\bf e}}

\newcommand{\hv}{{\bf h}}

\newcommand{\uv}{{\bf u}}

\newcommand{\vv}{{\bf v}}
\newcommand{\xv}{{\bf x}}

\newcommand{\Em}{{\bf E}}

\newcommand{\Gm}{{\bf G}}
\newcommand{\Hm}{{\bf H}}

\newcommand{\Wm}{{\bf W}}
\newcommand{\Vm}{{\bf V}}

\newcommand{\Ym}{{\bf Y}}
\newcommand{\Zm}{{\bf Z}}

\newcommand{\Cc}{{\cal C}}

\newcommand{\Lc}{{\cal L}}

\newcommand{\Nc}{{\cal N}}

\newcommand{\Pc}{{\cal P}}

\newcommand{\Sc}{{\cal S}}

\newcommand{\define}{\stackrel{\triangle}{=}}

\newcommand{\eq}[1]{(\ref{#1})}
\newcommand{\her}{{\sf H}}

\def\BibTeX{{\rm B\kern-.05em{\sc i\kern-.025em b}\kern-.08em
    T\kern-.1667em\lower.7ex\hbox{E}\kern-.125emX}}

\begin{document}

\title{Quantized vs. Analog Feedback for the MIMO Downlink: A
Comparison between Zero-Forcing Based Achievable Rates}


\author{\authorblockN{Giuseppe Caire}
\authorblockA{University of Southern California\\
Los Angeles CA, 90089 USA\\
}
\and
\authorblockN{Nihar Jindal}
\authorblockA{University of Minnesota \\
Minneapolis MN, 55455 USA\\
}
\and
\authorblockN{Mari Kobayashi}
\authorblockA{CTTC \\
Barcelona, Spain\\
} \and
\authorblockN{Niranjay Ravindran}
\authorblockA{University of Minnesota \\
Minneapolis MN, 55455 USA\\
} } \maketitle \vspace{-1cm}
\begin{abstract}
We consider a MIMO fading broadcast channel and compare the
achievable ergodic rates when the channel state information at the
transmitter is provided by ``analog'' noisy feedback or by quantized
(digital) feedback. The superiority of digital feedback is shown,
with perfect or imperfect CSIR, whenever the number of feedback
channel uses per channel coefficient is larger than 1. Also, we show
that by proper design of the digital feedback link, errors in the
feedback have a minor effect even by using very simple uncoded
modulation. Finally, we show that analog feedback achieves a
fraction $1 - 2F$ of the optimal multiplexing gain even in the
presence of a feedback delay, when the fading belongs to the class
of ``Doppler processes'' with normalized maximum Doppler frequency
shift $0 \leq F < 1/2$.
\end{abstract}

\section{Model setup and background}

We consider a multi-input multi-output (MIMO) Gaussian broadcast channel modeling
the downlink of a system where the base station
(transmitter) has $M$ antennas and $K$ user terminals (receivers) have one antenna each.
A channel use of such channel is described by
\begin{equation} \label{model}
y_k = \hv_k^\her \xv + z_k, \;\; k = 1,\ldots,K
\end{equation}
where $y_k$ is the channel output at receiver $k$, $z_k \sim \Cc\Nc(0,N_0)$ is the corresponding
AWGN, $\hv_k \in \CC^M$ is the vector of channel coefficients from the $k$-th receiver to the transmitter antenna array
and $\xv$ is the channel input vector. The channel input is subject to the average power constraint
$\EE[|\xv|^2] \leq P$.

We assume that the channel {\em state}, given by the collection of
all channel vectors $\Hm = [\hv_1,\ldots,\hv_K] \in \CC^{M \times
K}$, varies in time according to a  block fading model where $\Hm$
is constant over each frame of length $T$ channel uses, and evolves
from frame to frame according to an ergodic stationary jointly
Gaussian process; i.i.d. block-fading channel, where the entries of
$\Hm$ are Gaussian i.i.d. with elements $\sim \Cc\Nc(0,1)$ is a
special case of this.

\subsection{Capacity results}

If $\Hm$ is perfectly and instantaneously known to all terminals
(perfect CSIT and CSIR), the capacity region of the channel
\eq{model} is obtained by MMSE-DFE beamforming and Gaussian
dirty-paper coding (see
\cite{caire2003atm,weingarten2004crg} and references therein).
Because of simplicity and robustness to non-perfect CSIT, simpler
{\em linear precoding} schemes with standard Gaussian coding have
been extensively considered. A particularly simple scheme consists
of zero-forcing (ZF) beamforming, where the transmit signal is
formed as $\xv = \Vm \uv$, such that $\Vm \in \CC^{M \times K}$ is a
zero-forcing beamforming matrix and $\uv \in \CC^K$ contains the
symbols from $K$ independently generated Gaussian codewords. For $K
\leq M$, the $k$-th column $\vv_k$ of $\Vm$ is chosen to be a unit
vector orthogonal to the subspace $\Sc_k = {\rm span}\{\hv_j : j
\neq k\}$.
In this case, the achievable sum rate is given by
\begin{equation} \label{RsumZF}
R^{\rm ZF}  = \max_{\sum_k \EEsmall[\Pc_k(\Hm)] \leq P} \;
\sum_{k=1}^K \EE\left [ \log \left (1 + \frac{|\hv_k^\her\vv_k|^2
\Pc_k(\Hm)}{N_0} \right ) \right ].
\end{equation}
We consider the situation where $K=M$, and thus do not consider user
selection.
Furthermore, we are mainly interested in the high-spectral
efficiency regime, where we can characterize the achievable sum rate
as $\kappa \log P/N_0 + O(1)$, and $\kappa$ is the ``system
multiplexing gain'' or ``pre-log factor'' of the ergodic sum rate.
Hence, it is well-known that using uniform power $\Pc_k = P/M$ for
all $k = 1,\ldots,M$, rather than performing optimal water-filling,
incurs a loss only in the $O(1)$ term, and we shall restrict to this
choice in the rest of this paper.

It is well-known that, under perfect CSIT and CSIR, both the optimal
``Dirty-Paper'' sum-rate $C$ and the zero-forcing sum-rate $R^{\rm
ZF}$ are equal to $M \log P/N_0 + O(1)$. On the contrary, under
non-perfect CSIT the rate sum may behave in a radically different
way; for example, if there is perfect CSIR and no CSIT when $\Hm$
has i.i.d. Gaussian entries, the sum rate is equal to $\log P/N_0 +
O(1)$ \cite{caire2003atm}

\subsection{Channel state feedback models}

We consider some specific CSIT and CSIR models and derive lower-bounds to the corresponding
achievable ergodic rates by analyzing a {\em naive} beamforming scheme that computes a mismatched ZF
beamforming matrix $\widehat{\Vm}$ from the CSIT.
In particular, we consider an ``analog'' CSIT feedback scheme where the transmitter observation at
frame time $t$ is given by
\begin{equation} \label{analog-feedback}
\{ \Gm(\tau) = \sqrt{\beta P} \Hm(\tau) + \Wm(\tau) : \tau = -\infty,\ldots,t-d\}
\end{equation}
where $\{\Wm(\tau)\}$ is a spatially and spectrally white Gaussian
process with elements $\sim \Cc\Nc(0,N_0)$ and $d$ is the feedback
delay. This models the case where the channel coefficients are
explicitly transmitted on the reverse link (uplink) using
unquantized quadrature-amplitude modulation
\cite{FastCSI,mari-jsac,motorola-guys,Mandayam}. The power scaling
$\beta$ corresponds to the number of channel uses per channel
coefficient, assuming that transmission in the feedback channel has
fixed peak power $P$ and that the channel state vector is modulated
by a $\beta M \times M$ unitary spreading matrix \cite{FastCSI}. A
simplifying assumption of this work is that we consider no fading
and orthogonal access in the CSIT feedback link, and we assume that
the SNR on the feedback channel is equivalent to the un-faded
downlink SNR ($P/N_0$).

A different CSIT feedback approach is based on quantizing the
channel vector at each receiver and transmitting back to the base
station a packet of $B$ bits, representing the corresponding
quantization index. If a random ensemble of quantization schemes is
used (referred to as {\em Random Vector Quantization}, or RVQ), in
\cite[Theorem 1]{Jindal} it is shown that the gap between ZF with
ideal CSI and the naive ZF scheme is given by
\begin{equation} \label{rate-gap-digital}
\Delta R_{\rm quant.} \leq \log\left (1 + \frac{P}{N_0}
2^{-\frac{B}{M-1}} \right ).
\end{equation}

\section{Rate gap bound for analog CSIT feedback} \label{analog.sect}

In the case of i.i.d. block fading and no feedback delay, the analog
CSIT feedback yields the observation of $\Gm = \sqrt{\beta P} \Hm +
\Wm$ at the beginning of every frame. The transmitter computes the
MMSE estimate of the channel matrix, $\widehat{\Hm} =
\frac{\sqrt{\beta P}}{N_0 + \beta P} \Gm$. The $k$-th column
$\widehat{\vv}_k$ of $\widehat{\Vm}$ is a unit vector orthogonal to
the subspace $\Sc_k = {\rm span}\{\widehat{\hv}_j : j \neq k\}$.
Notice that we can write $\Hm = \widehat{\Hm} + \Em$, where
$\widehat{\Hm}$ and $\Em$ are mutually independent and have Gaussian
i.i.d. components with mean zero and variance $\frac{\beta
P}{N_0\sigma_e^2}$ and $\sigma_e^2 = (1 + \beta P/N_0)^{-1}$,
respectively.

The signal at the $k$-th receiver is given by
\begin{equation} \label{receiver-k}
y_k = (\hv_k^\her \widehat{\vv}_k) u_k + \sum_{j\neq k} (\ev_k^\her \widehat{\vv}_j) u_j + z_k
\end{equation}
We assume that the frame duration is long enough such that some training scheme can be used in the downlink channel.
Training allows each receiver to estimate: 1) the useful signal coefficient, $a_k = (\hv_k^\her \widehat{\vv}_k)$
and 2) the variance of the interference plus noise $\zeta_k = \sum_{j\neq k} (\ev_k^\her \widehat{\vv}_j) u_j + z_k$, given by
$\Sigma_k = \EE\left [ |\zeta_k|^2 |  \ev_k,\widehat{\Hm} \right ] =
N_0 + \sum_{j\neq k} |\ev_k^\her \widehat{\vv}_j|^2 P/M$.
This conditioning is due to the fact that $\Sigma_k$ is estimated on each frame,
and the coefficients $(\ev_k^\her \widehat{\vv}_j)$ are constant over each frame and change from frame
to frame, following the block i.i.d. fading model.
The maximum achievable rate of user $k$ subject to the above assumptions is lowerbounded by
assuming a Gaussian input $u_k = u_k^G \sim \Cc\Nc(0,P/M)$,
and by considering the worst-case noise plus interference distribution in every frame.
Using stationarity and ergodicity, we have
\footnote{With some abuse of notation, the term in the second line
of \eq{funny-lower-bound1} have the following meaning:
\begin{eqnarray*}
& \EE \left [ \inf_{\zeta_k : \EEsmall[|\zeta_k|^2] \leq \Sigma_k} I(u_k^G; y_k |a_k,\Sigma_k) \right ] &  \\
& \equiv & \\
& \int \inf_{\zeta_k : \EEsmall[|\zeta_k|^2] \leq \sigma} I(u_k^G; y_k |a_k,\Sigma_k=\sigma) dF(\sigma) &
\end{eqnarray*}
where $F(\sigma)$ denotes the cdf of $\Sigma_k$.}
\begin{eqnarray} \label{funny-lower-bound1}
R_k & \geq & \EE \left [ \inf_{\zeta_k : \EEsmall[|\zeta_k|^2] \leq \Sigma_k} I(u_k^G; y_k |a_k,\Sigma_k) \right ] \nonumber \\
& \stackrel{\rm (a)}{=} & \EE\left [ \log\left (1 + \frac{|a_k|^2P}{\Sigma_k M} \right ) \right ]
\end{eqnarray}
where (a) follows from \cite{hassibi2002hrc}, noticing that
$a_k u_k^G$ and $\zeta_k$ are uncorrelated (even after conditioning on $a_k,\Sigma_k$).

Next, we shall bound the rate gap incurred by the naive ZF beamforming and analog feedback
with respect to the ZF beamforming with ideal CSIT. Denoting by $R_k^{\rm ZF}$ the rate of user $k$ with uniform (across users) and constant
(in time) power allocation $\Pc_k(\Hm) = P/M$ in (\ref{RsumZF}), we have
\begin{footnotesize}
\begin{eqnarray} \label{rate-gap-analog-upperbound}
\Delta R_{\rm analog} & \define & R_k^{\rm ZF} - R_k \nonumber \\
& \leq &  \EE \left[ \log \left(1 +  \frac{|\hv_k^\her \vv_k|^2  P}{N_0M} \right) \right]
- \EE \left[ \log \left(1 +  \frac{|a_k|^2 P }{\Sigma_k M} \right ) \right ] \nonumber \\
& = & \EE \left[ \log\left (1 + \frac{|\hv_k^\her \vv_k|^2 P}{N_0M} \right ) \right] \nonumber \\
& & - \EE \left[ \log\left (1 + \frac{\left (\sum_{j \neq k} |\ev_k^\her \widehat{\vv}_j|^2 + |a_k|^2\right ) P}{N_0M} \right ) \right] \nonumber  \\
& & + \EE \left[ \log \left (1 + \sum_{j \neq k} \frac{|\ev_k^\her \widehat{\vv}_j|^2  P}{N_0M} \right ) \right] \nonumber  \\
&\stackrel{\rm (a)}{\leq}& \EE \left[ \log\left ( 1 + \sum_{j \neq k} \frac{|\ev_k^\her \widehat{\vv}_j|^2  P}{N_0M} \right ) \right] \nonumber  \\
&\stackrel{\rm (b)}{\leq}& \log \left(1 + \frac{P}{N_0M} \sum_{j \neq k}
\EE[ |\ev_k^\her \widehat{\vv}_j|^2 ] \right) \nonumber \\
&\stackrel{\rm (c)}{=}& \log \left(1 + \frac{\sigma_e^2 P}{N_0} \frac{M-1}{M} \right),
\end{eqnarray}
\end{footnotesize}
where (a) follows from the fact that
$\sum_{j \neq k} |\ev_k^\her \widehat{\vv}_j|^2 + |a_k|^2$
stochastically dominates $|\hv_k^\her \vv_k|^2$
since $|a_k|^2$ and $|\hv_k^\her\vv_k|^2$ are identically distributed,
(b) follows from Jensen's inequality and the final expression (c) follows by noticing that
the $\widehat{\Vm}$ is a deterministic function of $\widehat{\Hm}$ and therefore it is independent of $\Em$.
Therefore, we can write $\EE[ |\ev_k^\her \widehat{\vv}_j|^2 ] = \EE[\widehat{\vv}_j^\her \EE[\ev_k\ev_k^\her] \widehat{\vv}_j ] =
\sigma_e^2 \EE[|\widehat{\vv}_j|^2] = \sigma_e^2$, since $\widehat{\vv}_j$ has unit norm by construction.

\section{Comparison with quantized CSIT feedback} \label{analog-digital-comp.sect}

In this section we compare analog and digital feedback under the
assumptions of perfect CSIR, no feedback errors, and no feedback
delay. Replacing the estimation error variance $\sigma_e^2 = (1 +
\beta P/N_0)^{-1}$ in \eq{rate-gap-analog-upperbound} and further
upper bounding we obtain:
\begin{eqnarray} \label{analog-gap-highsnr}
\Delta R_{\rm analog}
& \leq & \log \left(1 + \frac{1}{\beta} \right).
\end{eqnarray}

Let us now consider digital feedback over the same channel.
The rate gap obtained in
\cite[Theorem 1]{Jindal} and reported in \eq{rate-gap-digital} is
further upperbounded by $ \log (1 + (P/N_0) \cdot 2^{ - \frac{B}{M}})$.
Let us assume (very unrealistically) that the digital feedback link can operate
error-free and at capacity, i.e., it can reliably transmit $\log(1 + P/N_0)$ bits per symbol.
For the same number of feedback channel periods, $\beta M$, the number of feedback bits per mobile is
$B = \beta M \log_2 (1 + P/N_0)$. Replacing this into the rate gap bound, we obtain:
\begin{equation} \label{digital-gap-highsnr}
\Delta R_{\rm quant.} \leq \log \left (1 + \frac{P/N_0}{(1 + P/N_0)^\beta} \right ).
\end{equation}
If $\beta = 1$ the quantized and analog feedback achieve essentially the same rate gap of at most
1 b/s/Hz. However, if $\beta > 1$, unlike the analog feedback case,
the rate gap of the quantized feedback vanishes for $P/N_0 \rightarrow \infty$.
and digital is far superior to analog for $\beta > 1$.

This conclusion finds an appealing interpretation in the context of
rate-distortion theory.
It is well-known (see \cite{Gastpar} and references therein) that
analog transmission is an optimal strategy to send a Gaussian source
over a Gaussian channel with minimal end-to-end quadratic
distortion.
In our case, the source is the Gaussian channel vector $\hv_k$ and
the noisy channel is the feedback AWGN channel with SNR $P/N_0$.
Hence, the fact that analog feedback cannot be essentially
outperformed for $\beta = 1$ is expected. However, it is also
well-known that if the channel rate is larger than the source rate
(i.e., less than one Gaussian source symbol arrives per channel
symbol, which corresponds to $\beta > 1$ in our case), then analog
is strictly suboptimal as compared to separate source and channel
coding because the distortion with analog transmission scales as
$1/\beta$ whereas it decreases exponentially with $\beta$ (i.e.,
along the vector quantizer R-D curve) for digital transmission.

\section{Effects of Imperfect CSIR} \label{csir.sect}

We now consider the scenario where each receiver has only a noisy
estimate of its channel acquired via downlink training.
In order to allow for channel estimation, $\beta_1 M$ shared pilots
($\beta_1 \geq 1$ symbols per antenna) are transmitted. Each
receiver estimates its channel on the basis of $\Ym = \sqrt{\beta_1
P} \Hm + \Zm$, which yields (after MMSE estimation) Gaussian error
with variance $(1 + \beta_1 P/N_0)^{-1}$. Terminals feed back
channel information immediately after completion of this training
phase.
After the transmitter has chosen beamforming vectors on the basis of
the channel feedback, an additional round of downlink training is
performed to enable coherent detection and allow each terminal to
estimate its useful signal coefficient $a_k = \bf{h}_k^H
\hat{\bf{v}}_k$.  This can be accomplished in $\beta_2 M$ symbols by
transmitting along each of the beamforming vectors for $\beta_2$
symbols.  If MMSE estimation of $a_k$ is performed, we have $a_k =
\hat{a}_k + f_k$ where $f_k$ and $\hat{a}_k$ are independent complex
Gaussian's with variance $\sigma_f^2 = \frac{N_0}{N_0 + \beta_2P}$
and $1 - \sigma_f^2$, respectively.\footnote{Note that additional
training is required because terminals do not know the channels of
other terminals, and thus are not aware of the chosen beamforming
vectors.}

Under this set of assumptions, a lower bound to $I(u_k;y_k |
\hat{a}_k)$ can be derived using techniques similar to those in
\cite{Medard,hassibi-hochwald-03it}.  Using this lower bound and
some steps similar to those leading to
(\ref{rate-gap-analog-upperbound}), the following upper bound to the
rate gap can be reached at:
\begin{eqnarray*} \label{rate-gap-csir}
\Delta R & \leq &  \log_2 \left( 1 + \frac{P}{N_0M} \left(
\sigma_f^2 + (M-1) \EE [ |\hv_k^\her \widehat{\vv}_j|^2 ] \right),
\right)
\end{eqnarray*}
where the multi-user interference term $\EE [ |\hv_k^\her
\widehat{\vv}_j|^2 ]$ depends on the CSIT and thus on the channel
feedback ($\beta$) as well as the accuracy of the initial training
($\beta_1$). We again assume that $\beta M$ symbols are devoted to
channel feedback (per mobile). If analog feedback is used, we get an
upper bound of:
\begin{eqnarray} \label{analog-gap-highsnr-csir}
\Delta R_{\rm analog}& \leq & \log \left(1 + \frac{1}{\beta_1} +
\frac{1}{M \beta_2} + \frac{1}{\beta} \right)
\end{eqnarray}
In the case of digital feedback, under the assumption that $B =
\beta M \log_2 (1 + P/N_0)$ feedback bits per mobile are sent in an
error-free manner, we get:
\begin{equation} \label{digital-gap-highsnr-csir}
\Delta R_{\rm quant.} \leq \log \left (1 + \frac{1}{\beta_1} +
\frac{1}{M \beta_2} + \frac{P/N_0}{(1 + P/N_0)^\beta} \right ).
\end{equation}
Comparing (\ref{analog-gap-highsnr-csir}) and
(\ref{digital-gap-highsnr-csir}) we come to the same general
conclusions as in Section \ref{analog-digital-comp.sect}: if
$\beta=1$ then digital and analog are equivalent, but if $\beta > 1$
 digital is superior to analog because the effect of feedback noise
vanishes at high SNR for digital but does not do so for analog.

\begin{figure}
\begin{center}
\includegraphics[width=8cm,height=6cm]{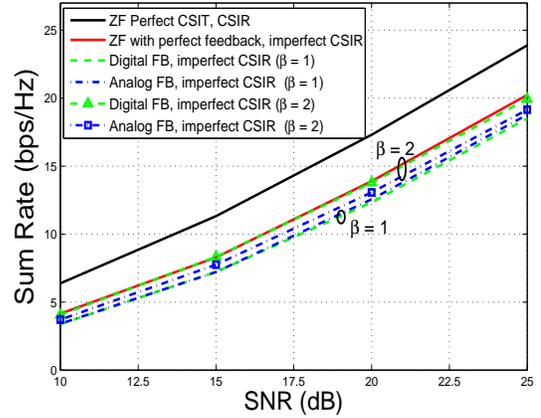}
\end{center}
\vspace{-0.5cm} \caption{Quantized vs. Analog Feedback with
Imperfect CSIR.} \label{fig-csir}
\end{figure}

There are, however, some important differences with the perfect CSIR
scenario.  First note that the imperfect CSIR leads to residual
interference that does not vanish with SNR; as a result, the rate
gap is not driven to $0$ even when $\beta>1$, assuming $\beta_1$ and
$\beta_2$ are fixed.  In addition, when $\beta_1 \approx \beta_2
\approx 1$, imperfect CSIR seems to have a considerably stronger
effect than feedback noise, thereby reducing the magnitude of
digital feedback's advantage. These effects are both visible in Fig.
\ref{fig-csir}, where analog and digital feedback curves are plotted
for $\beta_1=\beta_2=1$ and $\beta=1$ and $\beta=2$, along with the
throughput of an imperfect CSIR/perfect FB system.


Finally we comment on the tradeoff between downlink training
($\beta_1$) and channel feedback ($\beta$).  Since downlink pilots
are shared, training consumes only $\beta_1 M$ channel symbols.
Channel feedback, on the other hand, requires $\beta M$ channel
symbols \textit{per mobile}.  If the $M$ terminals can
simultaneously transmit on the feedback channel, perhaps utilizing
the $M$-antenna receive array at the base as described in
\cite{FastCSI}, then $\beta$ and $\beta_1$ are equivalent in terms
of system resources. For the case of analog feedback, from
(\ref{analog-gap-highsnr-csir}) we see that $\beta$ and $\beta_1$
should be chosen equal.  On the other hand, if digital feedback is
used, it is only necessary to choose $\beta > 1$ (so that the effect
of feedback noise vanishes), and the remainder of the resources
should be devoted to downlink training, i.e., to $\beta_1$.  This is
an additional advantage to digital whenever $\beta + \beta_1 > 2$.
Note that
there is also a tradeoff between $\beta_1$ and $\beta_2$, but that
the effect of the initial training ($\beta_1$) is considerably
stronger than the second phase.

\section{Effects of CSIT feedback errors} \label{errors.sect}

We now investigate the impact of removing the optimistic assumption
that the quantized feedback channel can operate error-free at
capacity.
We  consider a very simple CSIT feedback scheme that certainly
represents a lower bound on the best quantized feedback strategy.
The user terminals perform quantization using RVQ and transmit the
feedback bits using simple uncoded QAM. No intelligent mapping of
the quantization bits onto the QAM symbols is used, and therefore
even a single erroneous feedback bit from user $k$ results in CSIT
that is completely independent (due to the properties of RVQ) of the
actual $k$-th channel vector. Since uncoded QAM is used, error
detection is not possible and the base station computes beamforming
vectors based on the possibly erroneous feedback.

We again use $\beta M$ symbol periods to transmit the feedback bits.
There is a non-trivial tradeoff between quantization and channel
errors. In order to maintain a bounded gap, feedback must be scaled
at least as $(M-1) \log_2(1 + P/N_0) \approx M \log_2 P/N_0$.
Therefore, we consider sending $B = \alpha M \log_2 P/N_0$ for $1
\leq \alpha \leq \beta$ bits in $\beta M$ symbol periods, which
corresponds to $\frac{\alpha}{\beta} \log_2 (P/N_0)$ bits per QAM
 symbol.

From \cite{goldsmith2005wc}, using the fact that the QAM constellation size is equal to
$L = (P/N_0)^{\frac{\alpha}{\beta}}$, we have the following upper bound
to the symbol error probability for QAM modulation:
\begin{eqnarray} \label{qam-ser}
P_s
& \leq & 2 \exp \left (- \frac{3}{2} \left (\frac{P}{N_0} \right )^{1 - \alpha/\beta} \right )
\end{eqnarray}
For $\alpha = \beta$ (which means trying to signal at capacity with
uncoded modulation!) $P_s$ does not decreases with SNR and the
system performance is very poor. However, for $\alpha/\beta < 1$,
which corresponds to transmitting at a constant fraction of
capacity, $P_s \rightarrow 0$ as $P/N_0 \rightarrow \infty$. The
upper bound on the error probability of the whole quantized vector
(transmitted in $\beta M$ symbols) is given by $P_{e,fb} = 1 - (1 -
P_s)^{\beta M}$. A lower bound on the achievable ergodic rate is
obtained by assuming that when a feedback error occurs for user $k$
its SINR is zero
while if no feedback error occurs its rate is given $R_k^{\rm ZF} -
\Delta R_{\rm quant.}$, that is, the rate of ideal ZF decreased by
the (upper bound to) the rate gap. It follows that the ergodic rate
of user $k$
is upperbounded by
\begin{equation} \label{fb-errors}
R_k  \geq (1 - P_s)^{\beta M} \left ( R^{\rm ZF}_k -  \log\left (1 + (P/N_0)^{1 - \alpha} \right ) \right )
\end{equation}
Choosing $1 < \alpha < \beta$ we achieve both vanishing $P_s$ and vanishing $\Delta R_{\rm quant.}$ as $P/N_0 \rightarrow \infty$. Thus,
even under this very simple CSIT feedback scheme the optimal ZF performance can be eventually approached for sufficiently
high SNR.

\begin{figure}
\begin{center}
\includegraphics[width=8cm,height=6cm]{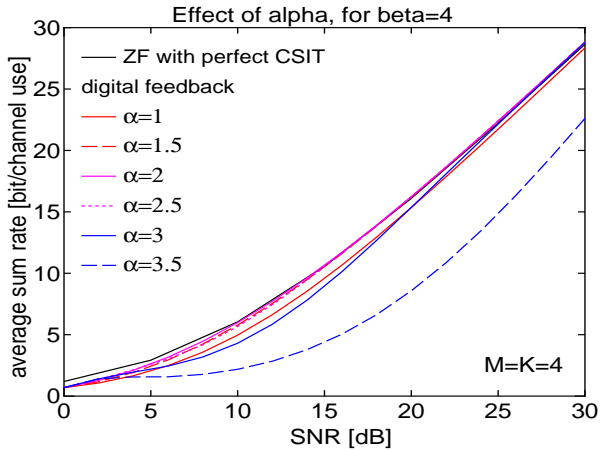}
\end{center}
\vspace{-0.5cm}
\caption{Quantized feedback with QAM modulation.}
\label{alpha}
\end{figure}

Fig.~\ref{alpha} shows the ergodic rate achieved by ZF beamforming
with quantized CSIT and QAM feedback transmission for $M = K = 4$,
independent Rayleigh fading, $\beta = 4$ and different values of
$\alpha$. It is noticed that by proper design of the feedback
parameters the performance can be made very close to the ideal CSIT
case.


\section{Effects of CSIT feedback delay} \label{feedback-delay.sect}

We consider now the case of analog feedback (assuming perfect CSIR)
when each entry of $\Hm$ evolves independently (in the block-fading
way described earlier) according to the same complex circularly
symmetric Gaussian stationary ergodic random process, denoted by
$\{h(t)\}$, with mean zero, variance 1 and power spectral density
(Doppler spectrum) denoted by $S_h(\xi)$, $\xi \in [-1/2,1/2]$.

Because of stationarity, without loss of generality we can focus on
$t=0$. We are interested in the linear MMSE estimation of $h(t)$
from the observation $\{g(\tau) : \tau = -\infty,t - d\}$ where,
following the analog feedback model \eq{analog-feedback}, we let
$g(\tau) = h(\tau) + w(\tau)$, with $w(\tau)$ i.i.d. $\sim
\Cc\Nc(0,\delta)$ and $\delta = \frac{N_0}{\beta P}$. In particular,
we consider the case of 1-step prediction ($d=1$) and the case of
filtering ($d=0$). From classical Wiener filtering theory
\cite{poor-book},
we have that the prediction error is given by
\begin{equation} \label{noisy-prediction}
\epsilon_1(\delta) = \exp\left ( \int_{-1/2}^{1/2} \log(\delta + S_h(\xi)) d\xi \right ) - \delta
\end{equation}
and that the filtering MMSE is given by
\begin{equation} \label{noisy-filtering}
\epsilon_0(\delta) = \frac{\delta \epsilon_1(\delta)}{\delta +
\epsilon_1(\delta)}.
\end{equation}

We shall discuss the rate gap bound \eq{rate-gap-analog-upperbound}
letting $\sigma_e^2 = \epsilon_d(N_0/(\beta P))$ for $d = 0,1$,
under different assumptions on the fading process $\{h(t)\}$.
We distinguish two cases: Doppler process and regular process. We
say that $\{h(t)\}$ is a Doppler process if $S_h(\xi)$ is strictly
band-limited in $[-F,F]$, where $F < 1/2$ is the maximum Doppler
frequency shift, given by $F = \frac{vf_c}{c} T_f$, where $v$ is the
mobile terminal speed (m/s), $f_c$ is the carrier frequency (Hz),
$c$ is light speed (m/s) and $T_f$ is the frame duration (s).
Furthermore, a Doppler process must satisfy $\int_{-F}^F \log
S_h(\xi) d\xi > -\infty$.
Following \cite{lapidoth2005acs}, we say that
$\{h(t)\}$ is a regular process if $\epsilon_1(0) > 0$. In
particular, a process satisfying the Paley-Wiener condition
\cite{poor-book} $\int_{-1/2}^{1/2} \log S_h(\xi) d\xi > -\infty$ is
regular.

A Doppler process satisfying our assumptions has prediction error
\begin{equation} \label{e1-doppler}
\epsilon_1(\delta) = \delta^{1 - 2F} \exp\left ( \int_{-F}^{F} \log(\delta + S_h(\xi)) d\xi \right ) - \delta
\end{equation}
{\bf No feedback delay ($d=0$).} In this case
\begin{eqnarray} \label{termP}
P \sigma_e^2
& = & \frac{N_0}{\beta} \frac{\epsilon_1\left (\frac{N_0}{\beta P}\right )}{\frac{N_0}{\beta P} + \epsilon_1\left (\frac{N_0}{\beta P}\right )}
\end{eqnarray}
Hence, $\lim_{P \rightarrow \infty} P\sigma_e^2 = \frac{N_0}{\beta}$ for both
Doppler and regular processes. For the latter, this is clear from the fact that $\epsilon_1(0) > 0$.
For the former, this follows from \eq{e1-doppler}. Applying
Jensen's inequality and the fact that $\int S_h(\xi) d\xi = 1$, we arrive at the upper bound
\begin{footnotesize}
\begin{equation} \label{termP-1}
\epsilon_1\left (\frac{N_0}{\beta P}\right ) \leq \left (\frac{N_0}{\beta P}\right )^{1-2F} \left [ \left (\frac{1}{2F} + \left (\frac{N_0}{\beta P}\right ) \right )^{2F} - \left (\frac{N_0}{\beta P}\right )^{2F} \right ]
\end{equation}
\end{footnotesize}
Using the fact that $\log$ is increasing, we arrive at the lower bound
\begin{footnotesize}
\begin{equation} \label{termP-2}
\epsilon_1\left (\frac{N_0}{\beta P}\right ) \geq \left (\frac{N_0}{\beta P}\right )^{1-2F} \left [ \exp\left (\int_{-F}^F \log S_h(\xi) d\xi\right ) - \left (\frac{N_0}{\beta P}\right )^{2F} \right ]
\end{equation}
\end{footnotesize}
These bounds yield that $\epsilon_1(N_0/\beta P) = \kappa P^{-(1-2F)} + O(1/P)$ for some constant $\kappa$.
Hence, $\epsilon_1 = O(P^{-(1-2F)})$ while $\delta = O(1/P)$, and the limits holds.

We conclude that in the case of no feedback delay the estimation
error is essentially dominated by the instantaneous observation and
not much improvement can be expected by taking into account the
channel memory if analog feedback is used.  With quantized feedback
the same may not be true because it is possible to exploit memory by
feeding back only the innovation process \cite{roh-rao04}; this is
under investigation.


{\bf Feedback delay ($d=1$).} In this case, the behavior of Doppler versus regular processes is radically different.
For Doppler processes, using \eq{termP-1} and \eq{termP-2},  we have that $P\sigma_e^2 = P\epsilon_1(N_0/\beta P) = \kappa P^{2F} + O(1)$.
It follows that the achievable rate sum is lowerbounded by
\begin{equation} \label{waccaz}
\sum_{k=1}^M R_k \geq M(1 - 2F) \log P + O(1)
\end{equation}
which implies a multiplexing gain of $M(1 - 2F)$.

\begin{figure}
\begin{center}
\includegraphics[width=8cm,height=6cm]{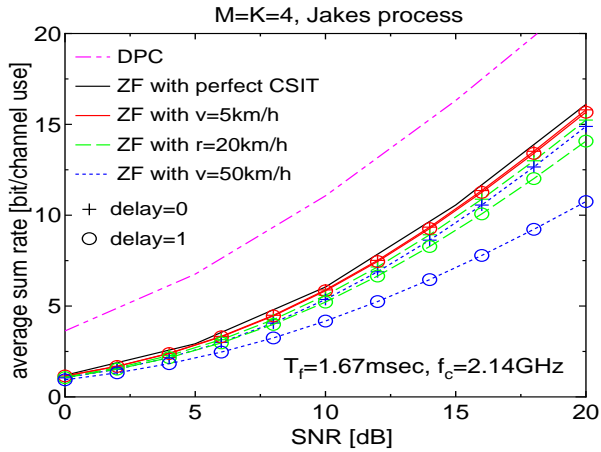}
\end{center}
\vspace{-0.5cm} \caption{Rates with feedback delay and Jakes'
correlation.} \label{jakes}
\end{figure}

For regular processes, on the contrary, we have that $P\sigma_e^2
\geq P \epsilon_1(0) = O(P)$. Hence, the rate gap grows like $\log
P$ and the achieved multiplexing gain is zero.  Furthermore, it can
be shown that the following is actually an upper bound to the
per-user rate, even when the feedback is noiseless:
\begin{eqnarray} \nonumber
R_k &\leq&  \log_2 \left( \frac{1}{1-r^2} + (M-1) \right) -
\frac{\psi(M)}{\log_e2} + \\ && \frac{1}{\log_e2} \left(
\frac{1}{2M-1} + \frac{1}{2M-2} \right)
\end{eqnarray}


In conclusions, the most noteworthy result of this analysis is that
under common fading models (Doppler processes), the analog feedback
scheme achieves a potentially high multiplexing gain even with
realistic, noisy and delayed feedback. Notice for example that with
mobile speed $v = 50$ km/h, $f_c = 2$ GHz, and frame duration $1$
ms, we have $F = 0.0926$. With $M = 4$ antennas we achieve a yet
respectable pre-log factor equal to $3.26$ instead of 4.\footnote{It
is interesting to notice here the parallel with the results of
\cite{lapidoth2005acs} on the high-SNR capacity of the single-user
scalar ergodic stationary fading channel with no CSIR and no CSIT,
where it is shown that for a class of {\em non-regular} processes
that includes the Doppler processes defined here, the high-SNR
capacity grows like $\Lc \log P$, where $\Lc$ is the Lebesgue
measure of the set $\{\xi \in [-1/2,1/2] : S_h(\xi) = 0\}$. In our
case, it is clear that $\Lc = 1 - 2F$.}

\begin{figure}
\begin{center}
\includegraphics[width=8cm,height=6cm]{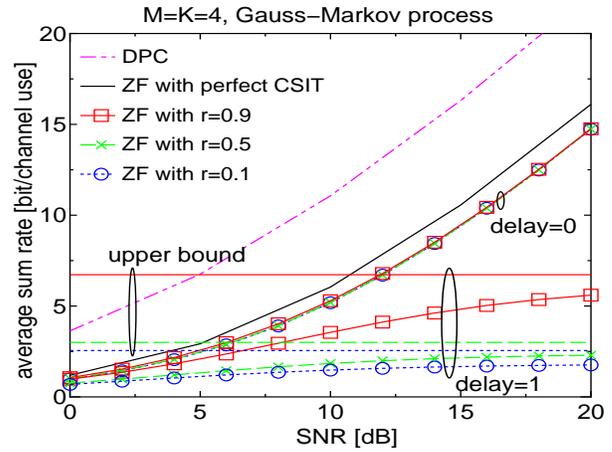}
\end{center}
\vspace{-0.5cm}
\caption{Rates with Gauss-Markov AR-1 correlation.}
\label{GMA}
\end{figure}

Figs.~\ref{jakes} and \ref{GMA} show the achievable ergodic rates for the Jakes' ``$J_0$'' correlation (strictly band-limited)
and the Gauss-Markov AR-1 correlation (regular process) for different first-lag correlation values.
For the AR-1 process with $d=1$ the system becomes interference limited.
On the contrary, the performance under Jakes' model degrades gracefully as the user mobility (Doppler bandwidth) increases.

\bibliographystyle{IEEEtran}
\bibliography{asilomar06}

\end{document}